\documentclass[runningheads]{llncs}
\usepackage{graphicx, amsfonts, amsmath, multirow, booktabs, subfigure}
\usepackage[misc]{ifsym} 
\newcommand{\reffig}[1]{Fig. \ref{#1}}
\usepackage{hyperref}
\begin{document}
%
\title{Automatically Segment the Left Atrium and Scars from LGE-MRIs Using a Boundary-focused nnU-Net}
\titlerunning{A Boundary-focused nnU-Net for the Left Atrium and Scars Segmentation}
%
\author{
Yuchen Zhang\inst{1} \and
Yanda Meng\inst{2} \and
Yalin Zheng\inst{2,3}$^{(\textrm{\Letter})}$  }
\authorrunning{Y. Zhang \textit{et al.}}
%
\institute{Center for Bioinformatics, Peking University, Beijing, China \and Department of Eye and Vision Science, University of Liverpool, Liverpool, UK \and Liverpool Centre for Cardiovascular Science, University of Liverpool and Liverpool Heart \& Chest Hospital, Liverpool, UK.\\ \email{yalin.zheng@liverpool.ac.uk}}
%

\maketitle              
\begin{abstract}
Atrial fibrillation (AF) is the most common cardiac arrhythmia. Accurate segmentation of the left atrial (LA) and LA scars can provide valuable information to predict treatment outcomes in AF. In this paper, we proposed to automatically segment LA cavity and quantify LA scars with late gadolinium enhancement Magnetic Resonance Imagings (LGE-MRIs). We adopted nnU-Net as the baseline model and exploited the importance of LA boundary characteristics with the TopK loss as the loss function.
Specifically, a focus on LA boundary pixels is achieved during training, which provides a more accurate boundary prediction.
On the other hand, 
a distance map transformation of the predicted LA boundary is regarded as an additional input for the LA scar prediction, which provides marginal constraint on scar locations. 
We further designed a novel uncertainty-aware module (UAM) to produce better results for predictions with high uncertainty.
Experiments on the LAScarQS 2022 dataset demonstrated our model's superior performance on the LA cavity and LA scar segmentation. Specifically, we achieved 88.98\% and 64.08\% Dice coefficient for LA cavity and scar segmentation, respectively. We will make our implementation code public available at \url{https://github.com/level6626/Boundary-focused-nnU-Net.}

\keywords{3D U-Net \and Segmentation \and Left atrium \and Boundary focused \and Distance map}
\end{abstract}
\section{Introduction}
Atrial fibrillation (AF) is the most common heart rhythm disturbance worldwide, affecting over 33 million people as of 2020 \cite{chung2020lifestyle}. 
The shape and distribution of LA scars due to AF ablation treatments are established indicators of treatment outcome and long term prognosis \cite{ranjan2011gaps}. Thus, the accurate segmentation of the LA region and scars in MRI images is essential for ablation planning and the post-operation care. In recent years, late gadolinium enhancement magnetic resonance imaging (LGE-MRI) has proved to be a promising tool for scar visualization and evaluation. In LGE-MRI, scar regions are enhanced with high intensity compared with healthy tissues nearby \cite{siebermair2017assessment}. However, manual annotation of the LA region and scars is a time-consuming and subjective task. Hence, developing an automatic segmentation algorithm for the LA region and scars in LGE-MRI images is vital. To this end, we developed an accurate LA region and scar segmentation framework with precise objects boundaries.

\subsubsection{Related Works}
The development of deep learning methods has recently led to great improvements in biomedical segmentation tasks \cite{meng2020regression,meng2020cnn,meng2023bilateral}. For example, the popular U-Net \cite{ronneberger2015u} backbone used a u-shaped architecture consisting of a contracting path and an expansive path to extract features from multi-scales and recovered them to precise localization. It was the most commonly used backbone for LGE-MRI LA Segmentation Challenge in MICCAI 2018 \cite{li2022medical}. Dozens of variations of U-Net were proposed to boost its performance, such as residual connections \cite{alom2019recurrent} and attention modules \cite{oktay2018attention}. Specifically, \textit{Fabian et al.} \cite{isensee2018nnu} believed that fine-tuning a plain U-Net is more worthwhile than adding various architecture modifications. They proposed nnU-Net \cite{isensee2018nnu}, which explores the inherent properties of datasets to achieve automatic parameter configuration. Their framework facilitates and enables data preprocessing, network architecture selection, network training, and predictions post-processing without the need of detailed domain knowledge.

The use of loss functions is of importance for segmentation tasks. For example, \textit{Cheng et al.} \cite{cheng2021boundary} proposed a boundary IoU metrics, which has been widely adopted by previous segmentation methods \cite{meng2021graph,meng2021bi,Meng_2022_Shape,meng2022dual}. \textit{Kervadec et al.} \cite{kervadec2019boundary} designed a novel integral way to compute the distance between two boundaries, avoiding differential computations of boundary locations. In the field of LA segmentation, 
\textit{Zhao et al.} \cite{zhao2021not} proposed a boundary loss on the distance between the predicted boundary and ground truth to optimize the segmentation results. \textit{Li et al.} \cite{li2022atrialjsqnet} employed a spatial encoding loss based on the distance probability map to introduce a regularization term for the LA segmentation.

On the other hand, the limited number of works \cite{yang2017fully,yang2020simultaneous,li2020atrial,li2022atrialjsqnet} reporting LA scar segmentation performance implies its challenging nature. The small size and discrete distribution of scars make it hard to achieve a high region-based evaluation score. In the related works \cite{li2020atrial,li2022atrialjsqnet,yang2017fully,yang2020simultaneous}, the predicted LA regions have been used to provide constraints for the scar segmentation to coerce the predicted scars located near the LA boundary. For example, \textit{Li et al.} \cite{li2022atrialjsqnet} designed a shape attention mechanism channeling the distance probability map of LA prediction to the scar predicting module, which is proved to be effective.

\begin{figure}\centering
\includegraphics[width=1\textwidth]{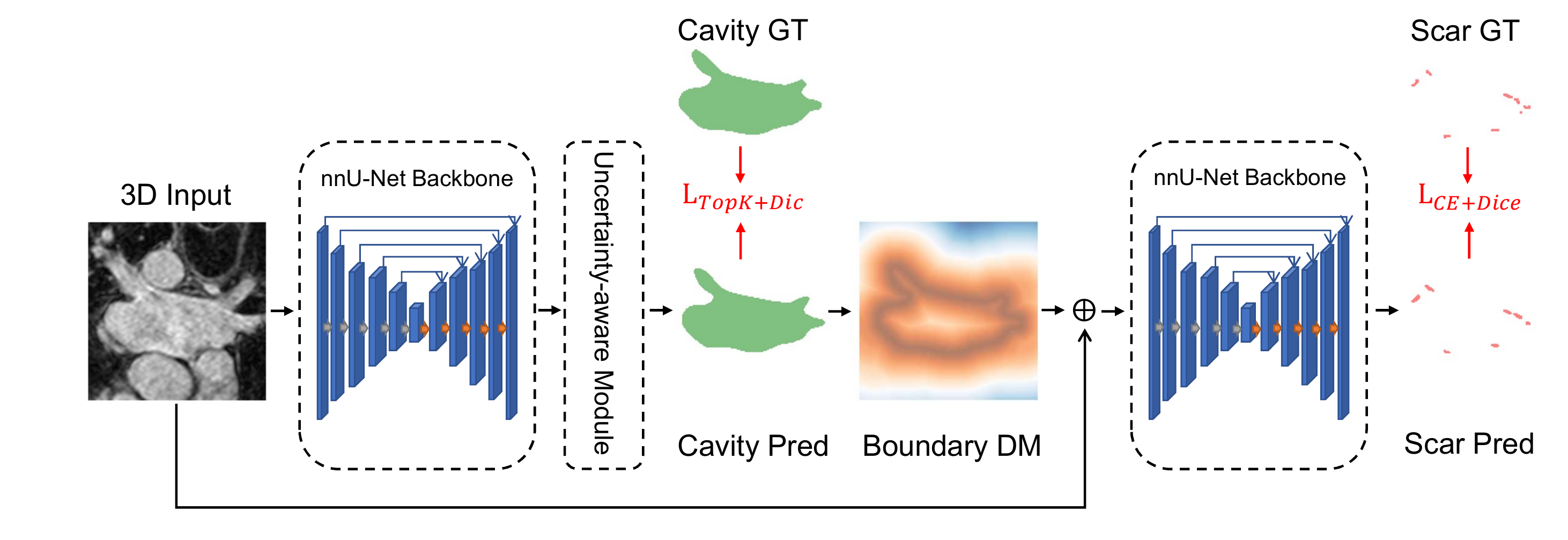}
\caption{Illustration of our proposed framework. GT means ground truth; Pred means prediction; DM means signed distance map; CE means cross entropy. The 3D input image is cropped for better visualization.} 
\label{fig:modified-nnUNet}
\end{figure}

\subsubsection{Our Contributions}
Inspired by the importance of the object boundary in the LA scar segmentation, we developed a boundary-based framework upon nnU-Net \cite{isensee2018nnu} for LAScarQS 2022 challenge \cite{li2021atrialgeneral,li2022atrialjsqnet,li2022medical}. Our framework consists of two stages: (1) For LA cavity segmentation, we adopted TopK \cite{wu2016bridging} in conjunction with Dice \cite{drozdzal2016importance} as the loss function, because TopK loss \cite{wu2016bridging} could automatically pay close attention to the boundary regions during the training process. (2) For LA scar quantification, we exploited the underlying spatial coherence between the LA cavities and the scars by directly concatenating the signed distance maps of the boundaries of the predicted LA cavities to the raw LGE-MRI images as the input. Notably, the outputs of the first stage were post-processed by our proposed novel uncertainty-aware module (UAM) to improve the final results of high-uncertainty predictions.

\section{Methods}

The framework of our method is shown in \reffig{fig:modified-nnUNet}. We adopted nnU-Net \cite{isensee2018nnu} as the backbone for segmenting both the LA cavities and scars. A combined loss function of TopK \cite{wu2016bridging} and Dice \cite{drozdzal2016importance} is adopted in the first stage. The predicted probability maps of the LA cavities are processed by UAM to achieve better results. After that, the predicted results of the LA cavities are transformed to a signed distance map of the LA boundaries. The inputs of the second stage are constructed by concatenating the raw LGE-MRI and the signed distance map of boundaries. Cross entropy and Dice \cite{drozdzal2016importance} loss are combined in the second stage.

\subsubsection{TopK loss for LA Segmentation}
Widely-used region-based losses, such as Dice \cite{drozdzal2016importance}, can usually lead to high accurate segmentations. However, it tends to overlook the sophisticated boundary shape because a large number of voxels inside the target shadow the significance of those on the boundary \cite{kervadec2019boundary,meng2021spatial}. This may lead to a relatively inaccurate LA boundary segmentation and in turn an inaccurate scar segmentation.
To address this, we adopted TopK loss \eqref{equ:topk} \cite{wu2016bridging} to introduce attention to the LA boundary during the training.
Boundary-focus methods \cite{li2020atrial,li2022atrialjsqnet} of LA segmentation attempt to give attention to the boundary. 
Actually, for objects that are not too small compared to the receptive field of CNN, the boundary is the most variable part of the prediction with the lowest certainty, the loss of boundary region is the highest among the prediction \cite{yang2018combating}. Based on the above assumption and reasoning, TopK loss is represented as:
\begin{equation}
L_{TopK}=-\frac{1}{N}\sum_{i\in K}g_i\log s_i
\label{equ:topk}
\end{equation}
where $g_{i}$ is the ground truth of voxel \textit{i}, $s_{i}$ is the corresponding predicted probability, and $K$ is the set of the \textit{ k\%} voxels with the lowest prediction accuracy.
While sole boundary-focused loss often causes training instability \cite{ma2021loss}, region-based loss, such as Dice loss \eqref{equ:dice} \cite{drozdzal2016importance}, is needed at the early stage of the training. We represent Dice loss as follow:
\begin{equation}
L_{Dice}=1-\frac{2|V_s\cap V_g|}{|V_s|+|V_g|}
\label{equ:dice}
\end{equation}
where $V_g$ is the ground truth label and $V_s$ is the prediction result of segmentation.
we coupled TopK with region-based Dice loss as our final loss function \eqref{equ:overall_loss} for the LA segmentation.
\begin{equation}
L=L_{TopK}+L_{Dice}
\label{equ:overall_loss}
\end{equation}

\subsubsection{Boundary Constraints for Scar Segmentation}

\begin{figure}[htbp] \centering
    \subfigure[]{
        \begin{minipage}[t]{0.29\linewidth}
        \label{subfig_boundaryGT}
        \centering
        \includegraphics[width=1.4in]{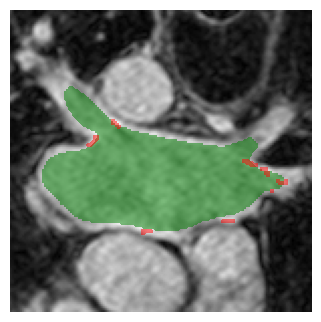}
        \end{minipage}
    }
    \subfigure[]{
        \begin{minipage}[t]{0.29\linewidth}
        \label{subfig_boundaryMask}
        \centering
        \includegraphics[width=1.4in]{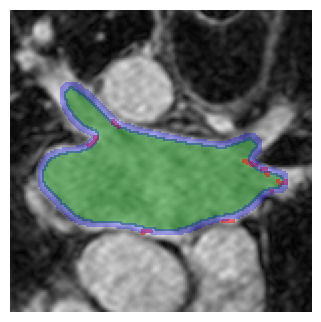}
        \end{minipage}
    }
    \subfigure[]{
        \begin{minipage}[t]{0.29\linewidth}
        \label{subfig_distmap}
        \centering
        \includegraphics[width=1.7in]{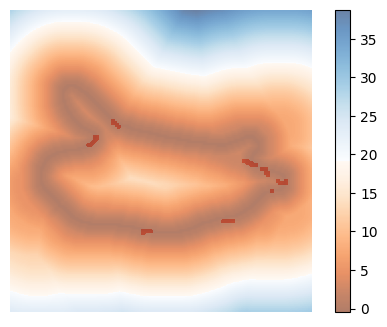}
        \end{minipage}
    }
    \caption{Boundary constraint: (a) overlap of the LA cavity label (green) and the scar label (red); (b) overlap of the the extracted boundary mask from the LA cavity ground truth (blue), LA cavity label (green) and the scar label (red); (c) overlap of the signed distance map of the boundary mask using Euclidean distance transformation and the scar label (red).}
    \label{fig:boundaryConstraint}
\end{figure}

Anatomically LA scars should be exactly located at the surface of the LA cavity. However, we found that LA scars were located in the adjacent area of the LA boundary. It is inaccurate to restrict the scars on the hard mask of the LA boundary. To address this, instead of a hard boundary mask, we adopted a soft boundary distance map to guide the prediction of LA scars.

To calculate the distance map of the LA boundary, we generated a mask of the boundary shown in \reffig{subfig_boundaryMask}. 
In detail, we substitute the max-pooling for the erosion operation as follows.
\begin{equation}
M_{b}=Pool_{d}(V_{g})+Pool_{e}(-V_{g})
\label{equ:boundary_mask}
\end{equation}
where $Pool_{d}$ and $Pool_{e}$ denote the 2D max pooling operation for mask dilation and erosion respectively. $Pool_{d}$ uses a kernel with size 5*5 and stride 1, while $Pool_{e}$ uses a kernel with size 3*3 and stride 1. 
The width of the boundary mask is 3 pixels, consisting of 2 pixels out of the exact boundary and 1 pixel inside. The results finely cover scar labels from the training data.

Given the boundary mask, the distance map of the LA boundary (shown in \reffig{subfig_distmap}) was calculated using Euclidean distance transformation as follows,
\begin{equation}
E(M)=[d(M_{ijk}, b_{ijk})_{ijk}]
\label{equ:eucl}
\end{equation}
\begin{equation}
D=E(-M) \cdot (-M) - (E(M) - 1) \cdot M
\label{equ:distmap}
\end{equation}
Where $d(\cdot)$ calculates the Euclidean distance between two voxels; $M_{ijk}$ denotes the voxels on the input mask; $b_{ijk}$ denotes the background voxel with the smallest Euclidean distance to the corresponding input point; $n$ is the number of dimensions. The original voxel spacing is taken into account in the transformation instead of assuming equal spacing along axes. The masked distance map of the boundary is then subtracted from its negated counterpart in \ref{equ:distmap}, giving the final signed distance map.
The signed distance map will be concatenated to the corresponding raw LGE-MRI image as the input of the network of the second stage.

\subsubsection{Highly Uncertain Prediction}
To boost the robustness of our framework, we designed an uncertainty-aware module (UAM) to detect the highly uncertain predictions. For these predictions, automatically lowering the threshold of the probability maps to the final mask outputs proves to be effective in improving the prediction results.
After training each fold of the five-fold cross-validation, we computed the sum of Shannon entropy \cite{shannon2001mathematical} for the output probability of each validation case. The mean and standard deviation was further calculated for all the cases after the training of all the folds is completed. When doing inference, the Shannon entropy \cite{shannon2001mathematical} of the output probability is compared to the population mean and deviation. We defined an outlier as three standard deviations away from the population mean. For outliers, the threshold of probability is lower to 0.2 rather than 0.5 to confirm a voxel as foreground.

\section{Experiments}
\subsubsection{Dataset and Preprocessing}
The public dataset used in this study is from the MICCAI 2022 Left Atrial and Scar Quantification \& Segmentation Challenge \cite{li2021atrialgeneral,li2022atrialjsqnet,li2022medical}. Task 1 ``LA Scar Quantification'' provides 60 post-ablation LGE-MRI training data with manual segmentation of LA and LA scars. Task 2 ``Left Atrial Segmentation from Multi-Center LGE MRIs'' provides 100 LGE-MRI training data with manual segmentations of the LA from three medical centers. Both pre-ablation and post-ablation images were included in this task. Images in the training dataset have two different sizes: 576*576*44 voxels and 640*640*44 voxels but with the same voxel dimension of 0.625*0.625*2.5 $mm^3$. We used the Task 2 dataset only for the LA segmentation whilst used the Task 1 dataset for the joint segmentation of LA and LA scars. For testing, Task 1 provides 10 LGE-MRI images and Task 2 provides 20 LGE-MRI images. Images in the testing dataset have two different sizes: 576*576*88 voxels and 640*640*88 voxels with the same voxel dimension of 1.0*1.0*1.0 $mm^3$.
All the input images were normalized by subtracting their mean and dividing by their standard deviation. Then, the input images were resampled by third-order spline interpolation and labels were resampled by one-order spline interpolation. Data augmentation was performed with the batchgenerators module, including Gaussian noise, gamma correction, random scaling, random rotations, and mirroring.

\subsubsection{Implementation Details}
For the baseline, We implemented the original nnU-Net \cite{isensee2018nnu} for the LA cavity and scar segmentation using Dice \cite{drozdzal2016importance} and cross-entropy as the loss function. All the inputs are original LGE-MRI images. We used stochastic gradient descent (SGD) with an initial learning rate of 0.01 and a momentum of 0.99 as the default settings. For the LA segmentation and LA scar segmentation on the Task 1 dataset, we ran training for 500 epochs and 130 epochs, respectively. For the LA segmentation on the Task 2 dataset, we ran training for 1000 epochs. Each epoch consists of 250 iterations. The learning rate was decayed in a polynomial style. If the average of the training loss does not improve during the previous 30 epochs, the learning rate will be divided by 5. No further uncertainty postprocessing was performed.

We implemented our framework in PyTorch with the same optimizer, learning rate scheduler and maximum epochs as the baseline. It takes 30 seconds per image to calculate the distance map of the LA boundary. We proposed the UAM only for LA segmentation. Because the scar segmentation is inherent uncertainty \cite{li2021atrialgeneral}, we set the softmax threshold for scar segmentation as 0.2 for all cases. We used 5 NVIDIA GeForce GTX3090 GPUs to train all the 5 folds with a batch size of 2. On the Task 1 dataset, it took 9.5 hours to train the LA segmentation model and another 2.5 hours to train the LA scar segmentation model. On the Task 2 dataset, it took 21.5 hours to train the LA segmentation model.

\section{Results}
\subsubsection{Comparison between Models}
The performance of our framework is compared to the baseline (original nnU-Net \cite{isensee2018nnu}). For the LA segmentation, Dice coefficient (Dice \cite{drozdzal2016importance}), Hausdorff distance (HD), and Average surface distance (ASD) were used to evaluate the results. For the LA scar segmentation, the Dice was used to evaluate the segmentation performance.

\begin{table}[htbp]\centering
\caption{LA segmentation with different \textit{K} values.  When $K=100$, TopK is the same as the cross-entropy. When $K=5$, it appears that the network cannot be trained because 5\% area is relative small for the network to learn the general region leading to the highly unstable training process.}\label{cavityResults}
\begin{tabular}{ccccccc}
\hline
K                        & \multicolumn{2}{c}{Dice (\%)} & \multicolumn{2}{c}{HD} &   \multicolumn{2}{c}{ASD}    \\ \cline{2-7} 
& Mean   & Std   & Mean  & Std   &  Mean   & Std    \\ \hline
100 & 88.78 & 5.72  & 16.94 & 5.27 & 1.749 & 0.804   \\ 
20 & 88.87  & 5.66  & 17.05 & 5.23 & 1.735 & 0.796   \\
10 & \textbf{88.96}  & 5.60 & \textbf{16.45} & 5.16 & \textbf{1.715} & 0.784  \\
5 & -  & - & - & - & - & -   \\ \hline
\end{tabular}
\end{table}

\begin{figure} [t]
    \centering
    \subfigure[Original image]{
        \begin{minipage}[t]{0.3\linewidth}
        \centering
        \includegraphics[width=1in]{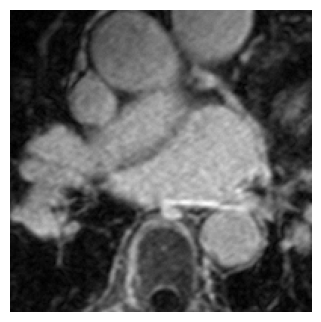}
        \end{minipage}
    }
    \subfigure[Cavity baseline]{
        \begin{minipage}[t]{0.3\linewidth}
        \centering
        \includegraphics[width=1in]{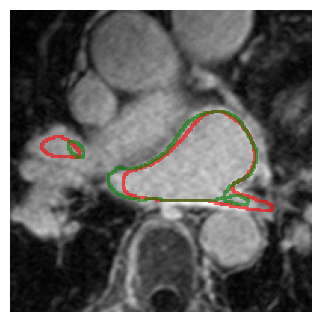}
        \end{minipage}
    }
    \subfigure[With TopK]{
        \begin{minipage}[t]{0.3\linewidth}
        \centering
        \includegraphics[width=1in]{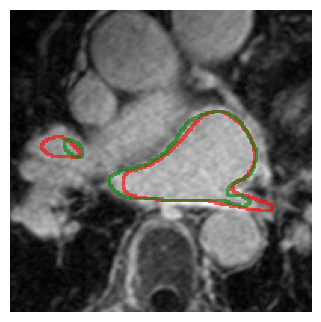}
        \end{minipage}
    }
    \subfigure[Original image]{
        \begin{minipage}[t]{0.3\linewidth}
        \centering
        \includegraphics[width=1in]{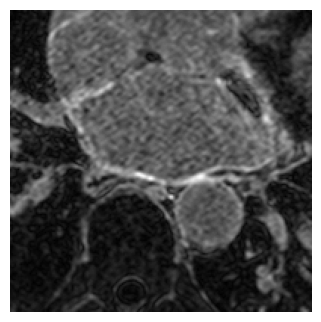}
        \end{minipage}
    }
    \subfigure[Scar baseline]{
        \begin{minipage}[t]{0.3\linewidth}
        \centering
        \includegraphics[width=1in]{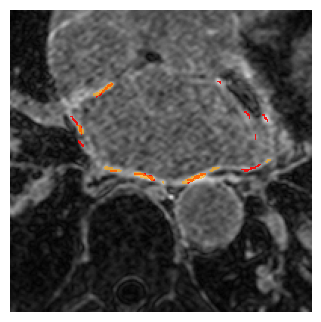}
        \end{minipage}
    }
    \subfigure[With distance map]{
        \begin{minipage}[t]{0.3\linewidth}
        \centering
        \includegraphics[width=1in]{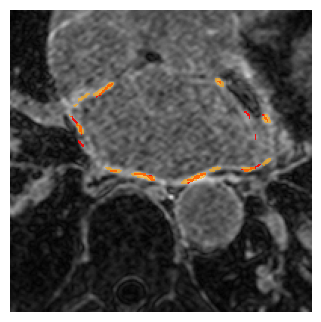}
        \end{minipage}
    }
    
    \caption{Visualisation of Segmentation Results. (a)-(c) show the segmentation results of the LA cavity, (d)-(f) show the segmentation results of the LA scars. The ground truth of the LA cavity is labelled with a red line, while the segmentation boundary from each model is in green. The ground truth of the LA scars is labelled in red, while the segmentation result from each model is in yellow. }
    \label{fig:seg_results}
    
\end{figure}

We conducted thorough experiments to validate the effectiveness of the value \textit{K} in the TopK loss.  
The testing results with different \textit{K} values for the LA segmentation are shown in Table \ref{cavityResults}. When $K=10$, substitution of TopK for cross-entropy reduces Hausdorff distance (HD) by 3.5\% and Average surface distance by 2.5\%, while Dice score is improved slightly by 0.2\%. We visualized the area of highest 10\% cross-entropy loss, i.e., TopK ($K=10$) focused area, during the training process in Supplementary Fig. 1. At the initial training steps, TopK focused areas are rather scattered when the network is learning the overall region of the target. While as the training goes on, these areas become more confined to the boundary area of the LA.
When \textit{K} equals to 5, the training process is highly unstable, because the 5\% area is relative small for the network to learn the overall region. When \textit{K} becomes bigger, however, the boundary focusing ability is gradually lost as shown by larger ASD and HD values.

\begin{table}[htbp]\centering
\caption{Results of the LA cavity prediction in the ablation study of the LA scar segmentation. UAM denotes uncertainty-aware module.}
\label{cavity_scarResults}
\begin{tabular}{cccccccc}
\hline
\multicolumn{1}{c}{Method} & \multicolumn{2}{c}{cavity Dice (\%)} & \multicolumn{2}{c}{cavity HD} &   \multicolumn{2}{c}{cavity ASD} \\ \hline 
LA cavity & Mean   & Std   & Mean  & Std   &  Mean   & Std \\ \hline
U-Net & 85.77 & 18.47  & 50.74 & 65.57 & 2.201 & 2.607 \\ 
U-Net+TopK & 88.09  & 11.67  & 25.86 & 15.51 & 2.110 & 2.386 \\
U-Net+TopK+UAM & \textbf{90.51}  & 4.53 & \textbf{23.32} & 8.32 & \textbf{1.64} & 0.987 \\ \hline
\end{tabular}
\end{table}

\begin{table}[htbp]\centering
\caption{Results of the LA scar prediction in the ablation study of the LA scar segmentation. DM denotes distance map; UAM denotes uncertainty-aware module.}
\label{scar_scarResults}
\begin{tabular}{cccc}
\hline
\multicolumn{2}{c}{Method} & \multicolumn{2}{c}{scar Dice (\%)}  \\ \hline 
LA cavity & LA scar & Mean & Std\\ \hline
U-Net & U-Net & 60.46 & 18.32   \\ 
U-Net+TopK & U-Net+DM  & 61.45 & 15.59  \\
U-Net+TopK+UAM & U-Net+DM  & \textbf{64.08} & \textbf{13.40}  \\ \hline
\end{tabular}
\end{table}

The testing results of the joint LA and LA scar segmentation are shown in Table \ref{cavity_scarResults}, Table \ref{scar_scarResults} and \reffig{fig:seg_results}. We conducted ablation studies on the TopK loss function, distance map of boundary (DM), and UAM. There is a significant 4.7 \% improvement in Dice, 27 \textit{mm} reduction in HD, and 0.56 \textit{mm} reduction in ASD of LA cavity segmentation when the TopK loss function was applied and the uncertainty-aware module was in action.
In the scar segmentation, the distance map concatenated with TopK and UAM brings a 4\% improvement in Dice over our baseline. The reason might be that the additional LA boundary information helps to constrain and locate scar predictions. The improved LA cavity prediction for the uncertainty cases also provides more accurate location guidance for scar predictions.

\section{Conclusion}
In this paper, we proposed a nnU-Net based approach to segment the LA cavity and LA scars from LGE-MRI images. Given the importance of the shape characteristics of the LA, we substituted the TopK loss function for the default cross-entropy, which automatically focuses on the LA boundary. To take into account the LA boundary in the scar prediction, we proposed to include distance information by concatenating the distance map of the LA boundary to raw LGE-MRI images. An uncertainty-aware module was designed for post-processing prediction results of poor-quality LGE-MRI images. Our proposed method has been evaluated on the LAScarQS 2022 dataset and the results have demonstrated its high accuracy on the LA  and LA scar segmentation. In the future, Our proposed method can be used as a promising tool to support the managements of cardiovascular diseases.

%
%
\bibliographystyle{splncs04}
\bibliography{LVSeg.bib}

\begin{thebibliography}{10}
\providecommand{\url}[1]{\texttt{#1}}
\providecommand{\urlprefix}{URL }
\providecommand{\doi}[1]{https://doi.org/#1}

\bibitem{alom2019recurrent}
Alom, M.Z., Yakopcic, C., Hasan, M., Taha, T.M., Asari, V.K.: Recurrent
  residual {U-Net} for medical image segmentation. Journal of Medical Imaging
  \textbf{6}(1),  014006 (2019)

\bibitem{cheng2021boundary}
Cheng, B., Girshick, R., Doll{\'a}r, P., Berg, A.C., Kirillov, A.: Boundary
  {IoU}: Improving object-centric image segmentation evaluation. In:
  Proceedings of the IEEE/CVF Conference on Computer Vision and Pattern
  Recognition. pp. 15334--15342 (2021)

\bibitem{chung2020lifestyle}
Chung, M.K., Eckhardt, L.L., Chen, L.Y., Ahmed, H.M., Gopinathannair, R.,
  Joglar, J.A., Noseworthy, P.A., Pack, Q.R., Sanders, P., Trulock, K.M.,
  et~al.: Lifestyle and risk factor modification for reduction of atrial
  fibrillation: a scientific statement from the american heart association.
  Circulation  \textbf{141}(16),  e750--e772 (2020)

\bibitem{drozdzal2016importance}
Drozdzal, M., Vorontsov, E., Chartrand, G., Kadoury, S., Pal, C.: The
  importance of skip connections in biomedical image segmentation. In: Deep
  Learning and Data Labeling for Medical Applications, pp. 179--187. Springer
  (2016)

\bibitem{isensee2018nnu}
Isensee, F., Petersen, J., Klein, A., Zimmerer, D., Jaeger, P.F., Kohl, S.,
  Wasserthal, J., Koehler, G., Norajitra, T., Wirkert, S., et~al.: nnu-net:
  Self-adapting framework for u-net-based medical image segmentation. arXiv
  preprint arXiv:1809.10486  (2018)

\bibitem{kervadec2019boundary}
Kervadec, H., Bouchtiba, J., Desrosiers, C., Granger, E., Dolz, J., Ayed, I.B.:
  Boundary loss for highly unbalanced segmentation. In: International
  conference on medical imaging with deep learning. pp. 285--296. PMLR (2019)

\bibitem{li2020atrial}
Li, L., Wu, F., Yang, G., Xu, L., Wong, T., Mohiaddin, R., Firmin, D., Keegan,
  J., Zhuang, X.: Atrial scar quantification via multi-scale {CNN} in the
  graph-cuts framework. Medical Image Analysis  \textbf{60},  101595 (2020)

\bibitem{li2021atrialgeneral}
Li, L., Zimmer, V.A., Schnabel, J.A., Zhuang, X.: Atrialgeneral: Domain
  generalization for left atrial segmentation of multi-center {LGE MRIs}. In:
  International Conference on Medical Image Computing and Computer-Assisted
  Intervention. pp. 557--566. Springer (2021)

\bibitem{li2022atrialjsqnet}
Li, L., Zimmer, V.A., Schnabel, J.A., Zhuang, X.: Atrialjsqnet: A new framework
  for joint segmentation and quantification of left atrium and scars
  incorporating spatial and shape information. Medical Image Analysis
  \textbf{76},  102303 (2022)

\bibitem{li2022medical}
Li, L., Zimmer, V.A., Schnabel, J.A., Zhuang, X.: Medical image analysis on
  left atrial {LGE MRI} for atrial fibrillation studies: A review. Medical
  Image Analysis p. 102360 (2022)

\bibitem{ma2021loss}
Ma, J., Chen, J., Ng, M., Huang, R., Li, Y., Li, C., Yang, X., Martel, A.L.:
  Loss odyssey in medical image segmentation. Medical Image Analysis
  \textbf{71},  102035 (2021)

\bibitem{meng2023bilateral}
Meng, Y., Bridge, J., Addison, C., Wang, M., Merritt, C., Franks, S., Mackey,
  M., Messenger, S., Sun, R., Fitzmaurice, T., et~al.: Bilateral adaptive graph
  convolutional network on ct based covid-19 diagnosis with uncertainty-aware
  consensus-assisted multiple instance learning. Medical Image Analysis
  \textbf{84},  102722 (2023)

\bibitem{Meng_2022_Shape}
Meng, Y., Chen, X., Zhang, H., Zhao, Y., Gao, D., Hamill, B., Patri, G., Peto,
  T., Madhusudhan, S., Zheng, Y.: Shape-aware weakly/semi-supervised optic disc
  and cup segmentation with regional/marginal consistency. In: International
  Conference on Medical Image Computing and Computer-Assisted Intervention.
  Springer (2022)

\bibitem{meng2020regression}
Meng, Y., Meng, W., Gao, D., Zhao, Y., Yang, X., Huang, X., Zheng, Y.:
  Regression of instance boundary by aggregated {CNN and GCN}. In: European
  Conference on Computer Vision. pp. 190--207. Springer (2020)

\bibitem{meng2020cnn}
Meng, Y., Wei, M., Gao, D., Zhao, Y., Yang, X., Huang, X., Zheng, Y.: {CNN-GCN}
  aggregation enabled boundary regression for biomedical image segmentation.
  In: International Conference on Medical Image Computing and Computer-Assisted
  Intervention. pp. 352--362. Springer (2020)

\bibitem{meng2021bi}
Meng, Y., Zhang, H., Gao, D., Zhao, Y., Yang, X., Qian, X., Huang, X., Zheng,
  Y.: {BI-GCN}: Boundary-aware input-dependent graph convolution network for
  biomedical image segmentation. In: 32nd British Machine Vision Conference:
  BMVC 2021. British Machine Vision Association (2021)

\bibitem{meng2022dual}
Meng, Y., Zhang, H., Zhao, Y., Gao, D., Hamill, B., Patri, G., Peto, T.,
  Madhusudhan, S., Zheng, Y.: Dual consistency enabled weakly and
  semi-supervised optic disc and cup segmentation with dual adaptive graph
  convolutional networks. IEEE Transactions on Medical Imaging p. in press
  (2022)

\bibitem{meng2021spatial}
Meng, Y., Zhang, H., Zhao, Y., Yang, X., Qian, X., Huang, X., Zheng, Y.:
  Spatial uncertainty-aware semi-supervised crowd counting. In: Proceedings of
  the IEEE/CVF International Conference on Computer Vision. pp. 15549--15559
  (2021)

\bibitem{meng2021graph}
Meng, Y., Zhang, H., Zhao, Y., Yang, X., Qiao, Y., MacCormick, I.J., Huang, X.,
  Zheng, Y.: Graph-based region and boundary aggregation for biomedical image
  segmentation. IEEE Transactions on Medical Imaging  \textbf{41}(3),  690--701
  (2021)

\bibitem{oktay2018attention}
Oktay, O., Schlemper, J., Folgoc, L.L., Lee, M., Heinrich, M., Misawa, K.,
  Mori, K., McDonagh, S., Hammerla, N.Y., Kainz, B., et~al.: Attention {U-net}:
  Learning where to look for the pancreas. arXiv preprint arXiv:1804.03999
  (2018)

\bibitem{ranjan2011gaps}
Ranjan, R., Kato, R., Zviman, M.M., Dickfeld, T.M., Roguin, A., Berger, R.D.,
  Tomaselli, G.F., Halperin, H.R.: Gaps in the ablation line as a potential
  cause of recovery from electrical isolation and their visualization using
  {MRI}. Circulation: Arrhythmia and Electrophysiology  \textbf{4}(3),
  279--286 (2011)

\bibitem{ronneberger2015u}
Ronneberger, O., Fischer, P., Brox, T.: U-net: Convolutional networks for
  biomedical image segmentation. In: International Conference on Medical Image
  Computing and Computer-assisted Intervention. pp. 234--241. Springer (2015)

\bibitem{shannon2001mathematical}
Shannon, C.E.: A mathematical theory of communication. ACM SIGMOBILE Mobile
  Computing and Communications Review  \textbf{5}(1),  3--55 (2001)

\bibitem{siebermair2017assessment}
Siebermair, J., Kholmovski, E.G., Marrouche, N.: Assessment of left atrial
  fibrosis by late gadolinium enhancement magnetic resonance imaging:
  methodology and clinical implications. JACC: Clinical Electrophysiology
  \textbf{3}(8),  791--802 (2017)

\bibitem{wu2016bridging}
Wu, Z., Shen, C., Hengel, A.v.d.: Bridging category-level and instance-level
  semantic image segmentation. arXiv preprint arXiv:1605.06885  (2016)

\bibitem{yang2020simultaneous}
Yang, G., Chen, J., Gao, Z., Li, S., Ni, H., Angelini, E., Wong, T., Mohiaddin,
  R., Nyktari, E., Wage, R., et~al.: Simultaneous left atrium anatomy and scar
  segmentations via deep learning in multiview information with attention.
  Future Generation Computer Systems  \textbf{107},  215--228 (2020)

\bibitem{yang2017fully}
Yang, G., Zhuang, X., Khan, H., Haldar, S., Nyktari, E., Ye, X., Slabaugh, G.,
  Wong, T., Mohiaddin, R., Keegan, J., et~al.: A fully automatic deep learning
  method for atrial scarring segmentation from late gadolinium-enhanced mri
  images. In: 2017 IEEE 14th International Symposium on Biomedical Imaging
  (ISBI 2017). pp. 844--848. IEEE (2017)

\bibitem{yang2018combating}
Yang, X., Wang, N., Wang, Y., Wang, X., Nezafat, R., Ni, D., Heng, P.A.:
  Combating uncertainty with novel losses for automatic left atrium
  segmentation. In: International Workshop on Statistical Atlases and
  Computational Models of the Heart. pp. 246--254. Springer (2018)

\bibitem{zhao2021not}
Zhao, Z., Puybareau, E., Boutry, N., G{\'e}raud, T.: Do not treat boundaries
  and regions differently: An example on heart left atrial segmentation. In:
  2020 25th International Conference on Pattern Recognition (ICPR). pp.
  7447--7453. IEEE (2021)

\end{thebibliography}

\section{Appendix}

\subsection{TopK Focused Area}

\begin{figure}[htbp] \centering
    \subfigure[Epoch 100]{
        \begin{minipage}[t]{0.3\linewidth}
        \label{subfig_ep100}
        \centering
        \includegraphics[width=2in]{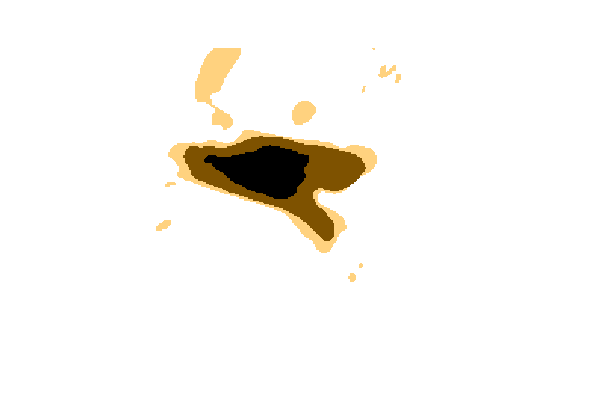}
        \end{minipage}
    }
    \subfigure[Epoch 200]{
        \begin{minipage}[t]{0.3\linewidth}
        \label{subfig_ep200}
        \centering
        \includegraphics[width=2in]{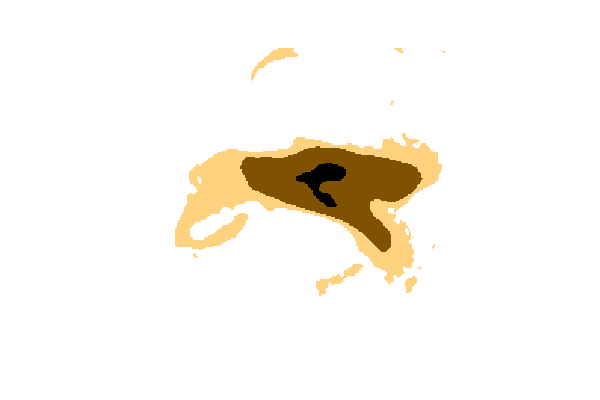}
        \end{minipage}
    }
    \subfigure[Epoch 250]{
        \begin{minipage}[t]{0.3\linewidth}
        \label{subfig_ep250}
        \centering
        \includegraphics[width=2in]{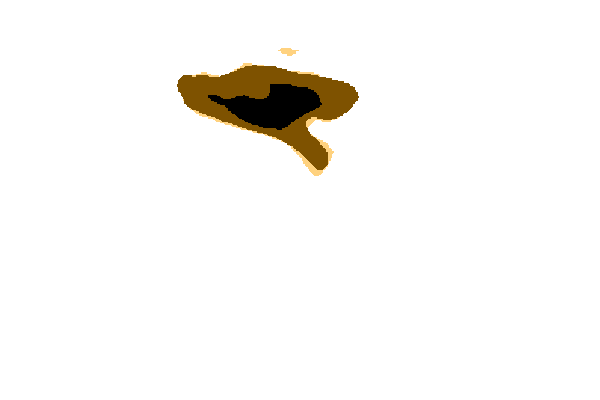}
        \end{minipage}
    }
    \subfigure[Epoch 350]{
        \begin{minipage}[t]{0.3\linewidth}
        \label{subfig_ep350}
        \centering
        \includegraphics[width=2in]{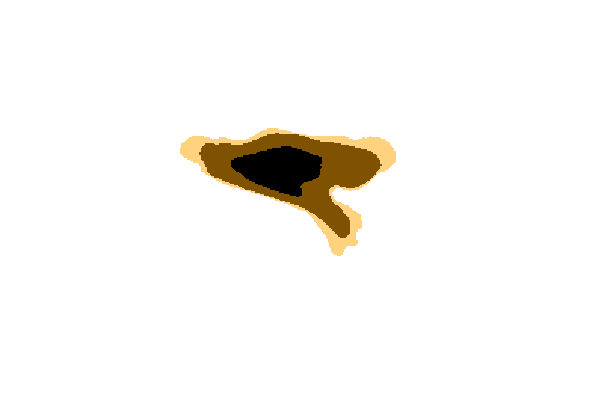}
        \end{minipage}
    }
    \subfigure[Epoch 450]{
        \begin{minipage}[t]{0.3\linewidth}
        \label{subfig_ep450}
        \centering
        \includegraphics[width=2in]{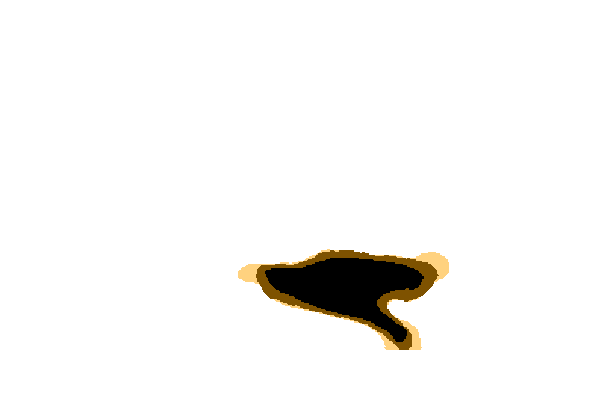}
        \end{minipage}
    }
    \subfigure[Epoch 500]{
        \begin{minipage}[t]{0.3\linewidth}
        \label{subfig_ep500}
        \centering
        \includegraphics[width=2in]{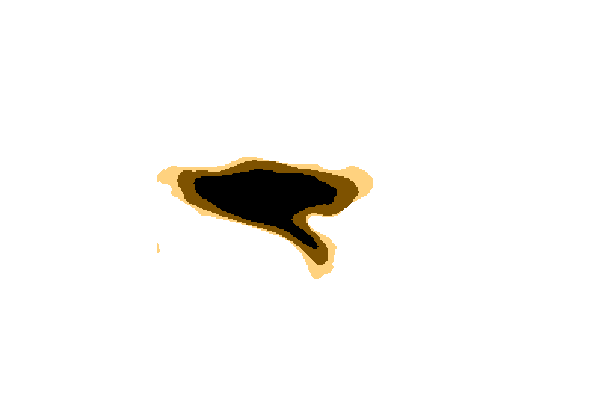}
        \end{minipage}
    }
    \caption{TopK Focused Area: Black + Brown: the ground truth of the LA cavity; Brown + Orange: the TopK focused area. Note, the position variations of the LA cavity between different epochs are caused by dynamic patch generation mechanism during the training process.}
    \label{fig:TopK}
\end{figure}

Here we provided the visualizations of the TopK-focused area \reffig{fig:TopK}. During the initial training steps, TopK-focused areas scatter because the network is learning the entire region of the target. While as the training continues, these areas become more localized to the boundary area of the LA.

\subsection{More Qualitative Results}

In \reffig{fig:seg_results}, we showed the effect of TopK loss function and uncertainty-aware module (UAM) on the segmentation of the LA cavities and the effect of the signed distance map (DM) of LA boundaries on the quantification of LA scars, respectively. Intuitively, TopK and UAM produce more faithful results on the LA cavity segmentation. The application of DM also improves the predictions of the LA scars.

\begin{figure}
    \centering
    \subfigure[Image]{
        \begin{minipage}[t]{0.17\linewidth}
        \centering
        \includegraphics[width=0.9in]{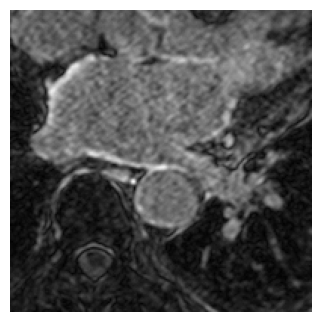}
        \includegraphics[width=0.9in]{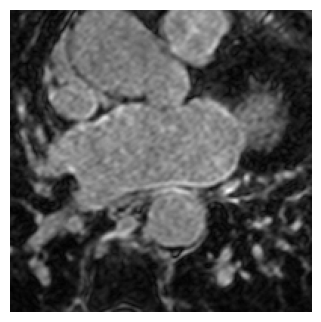}
        \includegraphics[width=0.9in]{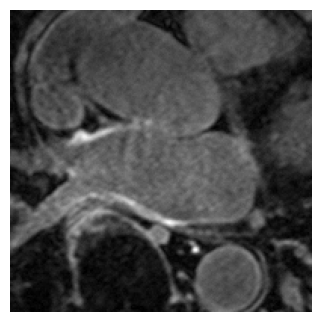}
        \end{minipage}
    }
    \subfigure[Cavity]{
        \begin{minipage}[t]{0.17\linewidth}
        \centering
        \includegraphics[width=0.9in]{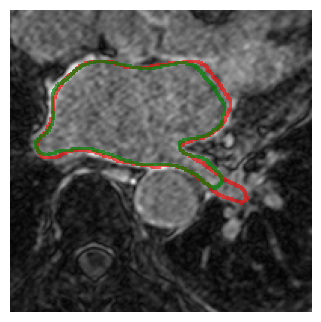}
        \includegraphics[width=0.9in]{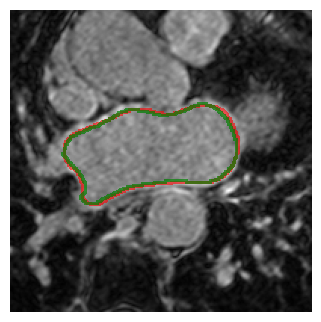}
        \includegraphics[width=0.9in]{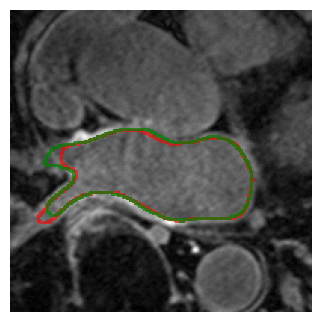}
        \end{minipage}
    }
    \subfigure[TopK+UAM]{
        \begin{minipage}[t]{0.18\linewidth}
        \centering
        \includegraphics[width=0.9in]{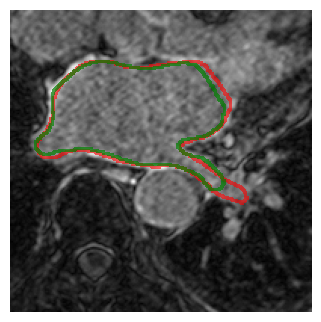}
        \includegraphics[width=0.9in]{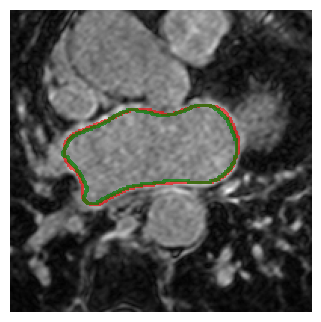}
        \includegraphics[width=0.9in]{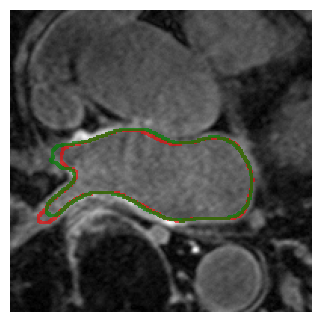}
        \end{minipage}
    }
    \subfigure[Scar]{
        \begin{minipage}[t]{0.17\linewidth}
        \centering
        \includegraphics[width=0.9in]{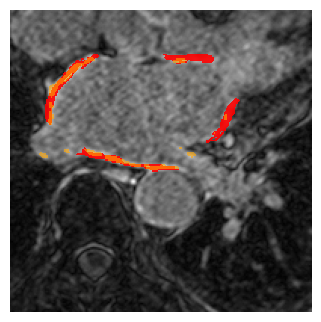}
        \includegraphics[width=0.9in]{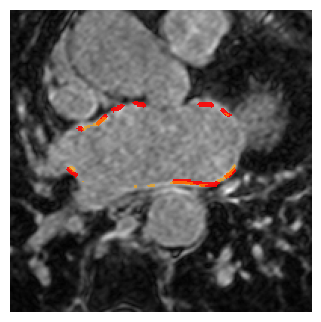}
        \includegraphics[width=0.9in]{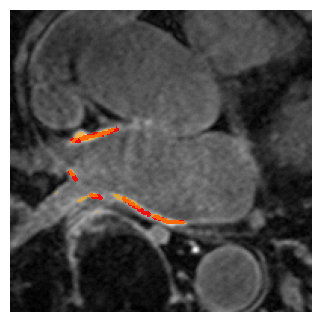}
        \end{minipage}
    }
    \subfigure[DM]{
        \begin{minipage}[t]{0.17\linewidth}
        \centering
        \includegraphics[width=0.9in]{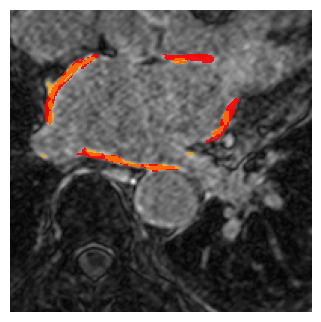}
        \includegraphics[width=0.9in]{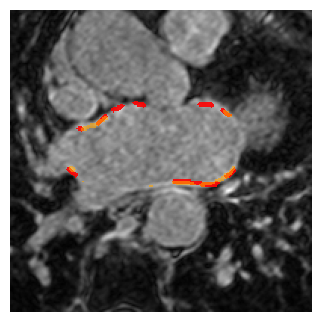}
        \includegraphics[width=0.9in]{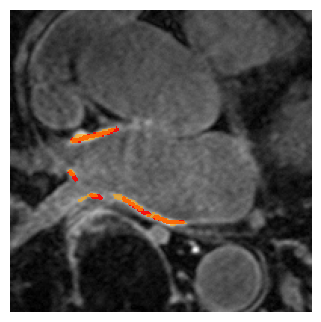}
        \end{minipage}
    }
    
    \caption{Visualization of Segmentation Results : (a) shows the input images, cropped for better visualization. (b)-(c) show the segmentation results of the LA cavities. (b) is from the baseline model. (c) shows the results when the TopK loss and the UAM are added to the baseline. (d)-(e) show the segmentation results of the LA scars. (d) is from the baseline model. (e) shows the results when the DM is used as an additional input. The ground truth of the LA cavity is labelled as a red line, while the segmentation boundary from each model is in green. The ground truth of the LA scars is labelled in red, while the segmentation results from each model is in yellow.}
    \label{fig:seg_results}
    
\end{figure}

\end{document}